\documentclass[prl,twocolumn,letterpaper, superscriptaddress]{revtex4-1}
\usepackage{graphicx}
\usepackage{dcolumn}
\usepackage{bm}
\usepackage{color}
\usepackage{tabularx}
\usepackage{array}
\usepackage{amsmath}
\usepackage{stmaryrd}  

\begin{document}

\title{Giant anisotropy of Gilbert damping in epitaxial CoFe films}

\author{Yi Li}
\affiliation{Department of Physics, Oakland University, Rochester, MI 48309, USA}
\affiliation{Materials Science Division, Argonne National Laboratory, Argonne, IL 60439, USA}

\author{Fanlong Zeng}
\affiliation{State Key Laboratory of Surface Physics, Department of Physics, Fudan University, Shanghai 200433, China}

\author{Steven S.-L. Zhang}
\affiliation{Materials Science Division, Argonne National Laboratory, Argonne, IL 60439, USA}

\author{Hyeondeok Shin}
\affiliation{Computational Sciences Division, Argonne National Laboratory, Argonne, IL 60439, USA}

\author{Hilal Saglam}
\affiliation{Materials Science Division, Argonne National Laboratory, Argonne, IL 60439, USA}
\affiliation{Department of Physics, Illinois Institute of Technology, Chicago IL 60616, USA}

\author{Vedat Karakas}
\affiliation{Materials Science Division, Argonne National Laboratory, Argonne, IL 60439, USA}
\affiliation{Department of Physics, Bogazici University, Bebek 34342, Istanbul, Turkey}

\author{Ozhan Ozatay}
\affiliation{Materials Science Division, Argonne National Laboratory, Argonne, IL 60439, USA}
\affiliation{Department of Physics, Bogazici University, Bebek 34342, Istanbul, Turkey}

\author{John E. Pearson}
\affiliation{Materials Science Division, Argonne National Laboratory, Argonne, IL 60439, USA}

\author{Olle G. Heinonen}
\affiliation{Materials Science Division, Argonne National Laboratory, Argonne, IL 60439, USA}

\author{Yizheng Wu}
\email{wuyizheng@fudan.edu.cn}
\affiliation{State Key Laboratory of Surface Physics, Department of Physics, Fudan University, Shanghai 200433, China}
\affiliation{Collaborative Innovation Center of Advanced Microstructures, Nanjing 210093, China}

\author{Axel Hoffmann}
\email{hoffmann@anl.gov}
\affiliation{Materials Science Division, Argonne National Laboratory, Argonne, IL 60439, USA}

\author{Wei Zhang}
\email{weizhang@oakland.edu}
\affiliation{Department of Physics, Oakland University, Rochester, MI 48309, USA}
\affiliation{Materials Science Division, Argonne National Laboratory, Argonne, IL 60439, USA}

\date{\today}

\begin{abstract}

Tailoring Gilbert damping of metallic ferromagnetic thin films is one of the central interests in spintronics applications. Here we report a giant Gilbert damping anisotropy in epitaxial Co$_{50}$Fe$_{50}$ thin film with a maximum-minimum damping ratio of 400 \%, determined by broadband spin-torque as well as inductive ferromagnetic resonance. We conclude that the origin of this damping anisotropy is the variation of the spin orbit coupling for different magnetization orientations in the cubic lattice, which is further corroborate from the magnitude of the anisotropic magnetoresistance in Co$_{50}$Fe$_{50}$.

\end{abstract}

\maketitle

\indent In magnetization dynamics the energy relaxation rate is quantified by the phenomenological Gilbert damping in the Landau-Lifshits-Gilbert equation \cite{GilbertIEEE2004}, which is a key parameter for emerging spintronics applications \cite{ManginNmat2006,KiselevNature2003,DussauxNcomm2010,ChumakNPhys2015,MironNmat2011}. Being able to design and control the Gilbert damping on demand is crucial for versatile spintronic device engineering and optimization. For example, lower damping enables more energy-efficient excitations, while larger damping allows faster relaxation to equilibrium and more favorable latency. Nevertheless, despite abundant approaches including interfacial damping enhancement \cite{UrbanPRL2001,mizukamiPRB2002,TserkovnyakRMP2005}, size effect \cite{nembachPRL2013,LiPRL2016} and materials engineering \cite{SchreiberSSC1995,OoganeJJAP2006,ScheckPRL2007}, there hasn't been much progress on how to manipulate damping within the same magnetic device. The only well-studied damping manipulation is by spin torque \cite{AndoPRL2008,WangAPL2011,LiuPRL2011,WeiZhangPRB2015}, which can even fully compensate the intrinsic damping \cite{DemidovNmat2012,HamadehPRL2014}. However the requirement of large current density narrows its applied potential. \\
\indent An alternative approach is to explore the intrinsic Gilbert damping anisotropy associated with the crystalline symmetry, where the damping can be continuously tuned via rotating the magnetization orientation. Although there are many theoretical predictions \cite{SteiaufPRB2005,BrataasPRL2008,FahnleJPD2008,GilmorePRB2010,VittoriaPRB2010}, most early studies of damping anisotropy are disguised by two-magnon scattering and linewidth broadening due to field-magnetization misalignment \cite{PlatowPRB1998,WoltersdorfPRB2004,LenzPRB2006,ZakeriPRB2007}. In addition, those reported effects are usually too weak to be considered in practical applications \cite{ChenNPhys2018,QinarXiv2018}.\\
\indent In this work, we show that a metallic ferromagnet can exhibit a giant Gilbert damping variation by a factor of four along with low minimum damping. We investigated epitaxial cobalt-iron alloys, which have demonstrated new potentials in spintronics due to their ultralow dampings \cite{SchoenNPhys2016,LeeNComm2018}. Using spin-torque-driven and inductive ferromagnetic resonance (FMR), we obtain a fourfold (cubic) damping anisotropy of 400\% in Co$_{50}$Fe$_{50}$ thin films between their easy and hard axes. For each angle, the full-range frequency dependence of FMR linewidths can be well reproduced by a single damping parameter $\alpha$. Furthermore, from first-principle calculations and temperature-dependent measurements, we argue that this giant damping anisotropy in Co$_{50}$Fe$_{50}$ is due to the variation of the spin-orbit coupling (SOC) in the cubic lattice, which differs from the anisotropic density of state found in ultrathin Fe film \cite{ChenNPhys2018}. We support our conclusion by comparing the Gilbert damping with the anisotropic magnetoresistance (AMR) signals. Our results reveal the key mechanism to engineer the Gilbert damping and may open a new pathway to develop novel functionality in spintronic devices.\\
\begin{figure}[htb]
 \centering
 \includegraphics[width=3.0 in]{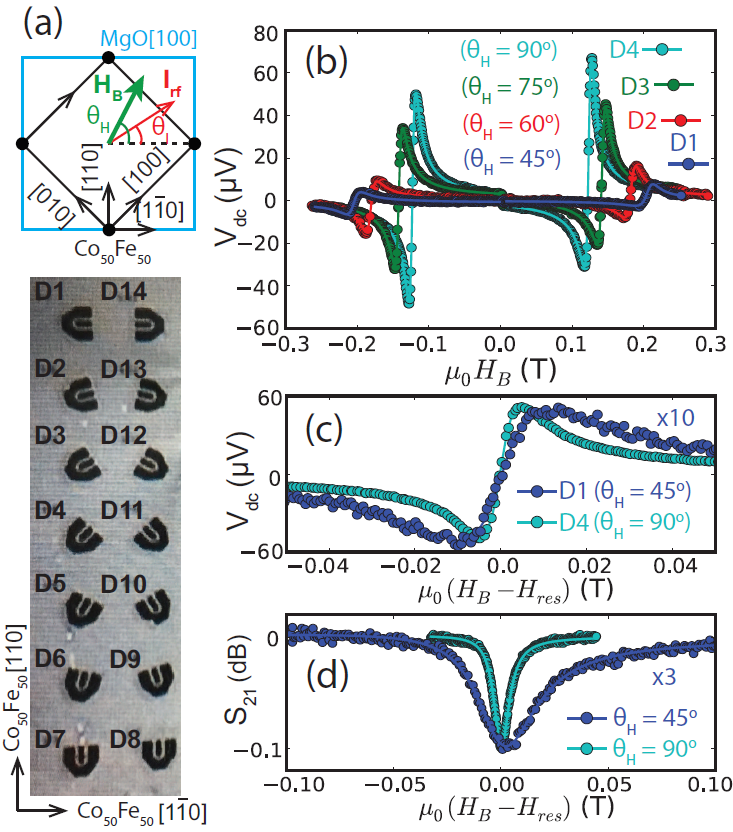}
 \caption{(a) Upper: crystalline structure, axes of bcc Co$_{50}$Fe$_{50}$ film on MgO(100) substrate and definition of $\theta_H$ and $\theta_I$. Lower: device orientation with respect to the CoFe crystal axis. (b) Spin-torque FMR lineshapes of i) CoFe(10 nm)/Pt devices D1 to D4 measured. (c) Resonances of D1 and D4 from (b) for $\mu_0H_{res}<0$. (d) Resonances of iii) CoFe(20 nm) for $\theta_H=45^\circ$ and $90^\circ$ measured by VNA FMR. In (b-d) $\omega/2\pi=20$ GHz and offset applies.}
 \label{fig1}
\end{figure}
\indent Co$_{50}$Fe$_{50}$ (CoFe) films were deposited on MgO(100) substrates by molecular beam epitaxy at room temperature, under a base pressure of 2$\times 10^{-10}$ Torr \cite{ChenAPL2011}. For spin-torque FMR measurements, i) CoFe(10 nm)/Pt(6 nm) and ii) CoFe(10 nm) samples were prepared. They were fabricated into 10 $\mu$m$\times$40 $\mu$m bars by photolithography and ion milling. Coplanar waveguides with 100-nm thick Au were subsequently fabricated \cite{WeiZhangPRB2015,JungfleischPRB2016}. For each layer structure, 14 devices with different orientations were fabricated, as shown in Fig. \ref{fig1}(a). The geometry defines the orientation of the microwave current, $\theta_I$, and the orientation of the biasing field, $\theta_H$, with respect to the MgO [100] axis (CoFe [1$\overline{1}$0]). $\theta_I$ ranges from 0$^\circ$ to 180$^\circ$ with a step of 15$^\circ$ (D1 to D14, with D7 and D8 pointing to the same direction). For each device we fix $\theta_H = \theta_I+45^\circ$ for maximal rectification signals. In addition, we also prepared iii) CoFe(20 nm) 40 $\mu$m$\times$200 $\mu$m bars along different orientations with transmission coplanar waveguides fabricated on top for vector network analyzer (VNA) measurements. See the Supplemental Materials for details \cite{supplement}.\\
\indent Fig. \ref{fig1}(b) shows the angular-dependent spin-torque FMR lineshapes of CoFe(10 nm)/Pt devices from different samples (D1 to D4, hard axis to easy axis) at $\omega/2\pi=20$ GHz. A strong magnetocrystalline anisotropy as well as a variation of resonance signals are observed. Moreover, the linewidth increases significantly from easy axis to hard axis, which is shown in Fig. \ref{fig1}(c). We have also conducted rotating-field measurements on a second CoFe(10 nm)/Pt device from a different deposition and the observations can be reproduced. This linewidth anisotropy is even more pronounced for the CoFe(20 nm) devices without Pt, measured by VNA FMR (Fig. \ref{fig1}d). For the CoFe(10 nm) devices, due to the absence of the Pt spin injector the spin-torque FMR signals are much weaker than CoFe/Pt and completely vanish when the microwave current is along the easy axes.\\
\begin{figure}[htb]
 \centering
 \includegraphics[width=3.2 in]{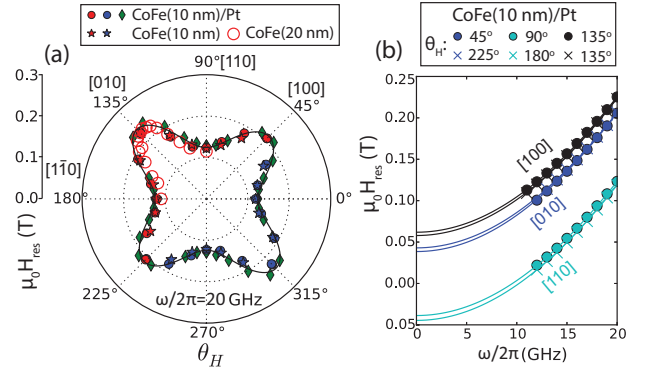}
 \caption{
 (a) Resonance field $\mu_0H_{res}$ as a function of $\theta_H$ at $\omega/2\pi=20$ GHz for different samples. Diamonds denote the rotating-field measurement from the second CoFe(10 nm)/Pt device. The black curve denotes the theoretical prediction. (b) $\mu_0H_{res}$ as a function of frequency for the CoFe(10 nm)/Pt devices. Solid curves denote the fits to the Kittel equation.}
 \label{fig2}
\end{figure}
\indent Figs. \ref{fig2}(a-b) show the angular and frequency dependence of the resonance field $H_{res}$. In Fig. \ref{fig2}(a), the $H_{res}$ for all four sample series match with each other, which demonstrates that the magnetocrystalline properties of CoFe(10 nm) samples are reproducible. A slightly smaller $H_{res}$ for CoFe(20 nm) is caused by a greater effective magnetization when the thickness increases. A clear fourfold symmetry is observed, which is indicative of the cubic lattice due to the body-center-cubic (bcc) texture of Co$_{50}$Fe$_{50}$ on MgO. We note that the directions of the hard axes has switched from [100] and [010] in iron-rich alloys \cite{LeeNComm2018} to [110] and [1$\overline{1}$0] in Co$_{50}$Fe$_{50}$, which is consistent with previous reports \cite{ShikadaJAP2009,KuschelJPD2012}. \\
\begin{figure}[htb]
 \centering
 \includegraphics[width=3 in]{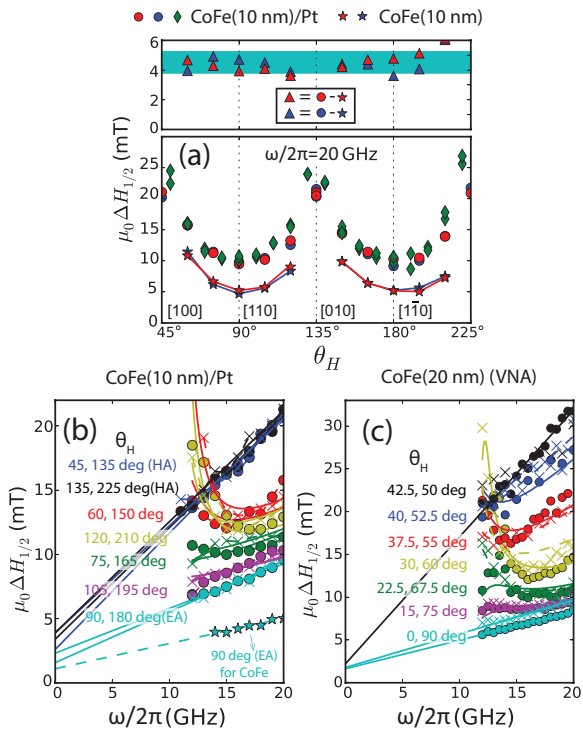}
 \caption{(a) $\mu_0\Delta H_{1/2}$ as a function of $\theta_H$ at $\omega/2\pi=20$ GHz for the CoFe(10 nm) series in Fig. \ref{fig2}(a). \textit{Top:} Addtional linewidth due to spin pumping of Pt. The green region denotes the additional linewidth as $4.5\pm0.7$ mT. (b-c) $\mu_0\Delta H_{1/2}$ as a function of frequency for (b) CoFe(10 nm)/Pt and (c) CoFe(20 nm) samples. Solid lines and curves are the fits to the data.}
 \label{fig3}
\end{figure}
\indent The magnetocrystalline anisotropy can be quantified from the frequency dependence of $\mu_0H_{res}$. Fig. \ref{fig2}(b) shows the results of CoFe(10 nm)/Pt when $H_B$ is aligned to the easy and hard axes. A small uniaxial anisotropy is found between [1$\overline{1}$0] ($0^\circ$ and $180^\circ$) and [110] ($90^\circ$) axes. By fitting the data to the Kittel equation $\omega^2/\gamma^2=\mu_0^2(H_{res}-H_k)(H_{res}-H_k+M_s)$, where $\gamma=2\pi(g_{eff}/2)\cdot 28$ GHz/T, we obtain $g_{eff}=2.16$, $\mu_0M_s=2.47$ T, $\mu_0H_k^{[100]}=40$ mT, $\mu_0H_k^{[010]}=65$ mT and $\mu_0H_k^{[110]}=\mu_0H_k^{[1\overline{1}0]}=-43$ mT. Taking the dispersion functions from cubic magnetocrystalline anisotropy \cite{FarleRPY1998,LiuPRB2003}, we obtain an in-plane cubic anisotropy field $\mu_0H_{4||}=48$ mT and a uniaxial anisotropy field $\mu_0H_{2||}=12$ mT. Fig. \ref{fig2}(a) shows the theoretical predictions from $H_{4||}$ and $H_{2||}$ in black curve, which aligns well with all 10-nm CoFe samples.\\
\indent With good magnetocrystalline properties, we now turn to the energy relaxation rate. Fig. \ref{fig3}(a) shows the full-width-half-maximum linewidths $\mu_0\Delta H_{1/2}$ of the spin-torque FMR signals at $\omega/2\pi=20$ GHz. Again, a fourfold symmetry is observed for CoFe(10 nm)/Pt and CoFe(10 nm), with the minimal (maximal) linewidth measured when the field lies along the easy (hard) axes. For CoFe(10 nm) devices, we did not measure any spin-torque FMR signal for $H_B$ along the hard axes ($\theta_H=45^\circ$, $135^\circ$ and $225^\circ$). This is due to the absence of the Pt spin injector as well as the near-zero AMR ratio when the rf current flows along the easy axes, which will be discussed later. For all other measurements, the linewidths of CoFe devices are smaller than for CoFe/Pt by the same constant, independent of orientation (upper diagram of Fig. \ref{fig3}a). This constant linewidth difference is due to the spin pumping contribution to damping from the additional Pt layer \cite{tserkovnyakPRL2002,GhoshAPL2011}. Thus we can deduce the intrinsic damping anisotropy from CoFe(10 nm)/Pt devices, with the damping shifted from CoFe(10 nm) devices by a constant and is much easier to measure. \\
\indent In Fig. \ref{fig3}(b-c) we show the frequency dependence of $\mu_0\Delta H_{1/2}$ of CoFe(10 nm)/Pt devices from spin-torque FMR and CoFe(20 nm) devices from VNA FMR. For both the easy and hard axes, linear relations are obtained, and the Gilbert damping $\alpha$ can be extracted from $\mu_0\Delta H_{1/2}=\mu_0\Delta H_0+2\alpha\omega/\gamma$ with the fits shown as solid lines. Here $\mu_0\Delta H_0$ is the inhomogeneous broadening due to the disorders in lattice structures. In Fig. \ref{fig3}(b) we also show the linewidths of the CoFe(10 nm) device along the easy axis ($\theta_H=90^\circ$), which has a significant lower linewidth slope than the easy axis of CoFe(10 nm)/Pt. Their differences yield a spin pumping damping contribution of $\Delta \alpha_{sp}=0.0024$. By using $\Delta \alpha_{sp}=\gamma\hbar g^{\uparrow\downarrow}/(4\pi M_st_M)$, we obtain a spin mixing conductance of $g^{\uparrow\downarrow}\text{(CoFe/Pt)}=25$ nm$^{-2}$, which is comparable to similar interfaces such as NiFe/Pt \cite{WeiZhangPRB2015sp,CaminalePRB2016}. For $\theta_H$ between the easy and hard axes, the low-frequency linewidth broadenings are caused by the deviation of magnetization from the biasing field direction, whereas at high frequencies the field is sufficient to saturate the magnetization. In order to find the damping anisotropy, we fit the linewidths to the angular model developed by Suhl \cite{SuhlPR1955,MizukamiJJAP2001}, using a single fit parameter of $\alpha$ and the extracted $H_{2||}$ and $H_{4||}$ from Fig. \ref{fig2}. The solid fitting curves in Fig. \ref{fig3}(b) nicely reproduce the experimental points. \\
\indent The obtained damping anisotropy for all the samples are summarized in Fig. \ref{fig4}, which is the main result of the paper. For CoFe(10 nm)/Pt samples, $\alpha$ varies from 0.0056 along the easy axis to 0.0146 along the hard axis. By subtracting the spin pumping $\Delta \alpha_{sp}$ from both values, we derive a damping anisotropy of 380\%. For CoFe(20 nm) samples measured by VNA FMR, $\alpha$ varies from 0.0054 to 0.0240, which yields an anisotropy of 440\% and reproduces the large anisotropy from spin-torque FMR. This giant damping anisotropy implies, technologically, nearly four times smaller critical current to switch the magnetization in a spin-torque magnetic random access memory, or to excite auto-oscillation in a spin-torque oscillator, by simply changing the magnetization orientation from the hard axis to the easy axis within the same device. In addition, we emphasize that our reported damping anisotropy is not subject to a dominant two-magnon scattering contribution, which would be manifested as a nonlinear linewidth softening at high frequencies \cite{LenzPRB2006,QinarXiv2018}. For this purpose we have extended the frequency of spin-torque FMR on CoFe(10 nm)/Pt up to 39 GHz, see the Supplemental Materials for details \cite{supplement}. We choose CoFe(10 nm)/Pt samples because they provide the best signals at high frequencies and the additional Pt layer significantly helps to excite the dynamics. Linear frequency dependence of linewidth persists throughout the frequency range and $\Delta H_0$ is unchanged for the two axes, with which we can exclude extrinsic effects to the linewidths. We also note that our result is substantially different from the recent report on damping anisotropy in Fe/GaAs \cite{ChenNPhys2018}, which is due to the interfacial SOC and disappears quickly as Fe becomes thicker. In comparison, the Gilbert damping anisotropy in Co$_{50}$Fe$_{50}$ is the intrinsic property of the material, is bonded to its bulk crystalline structure, and thus holds for different thicknesses in our experiments. \\
\begin{figure}[htb]
 \centering
 \includegraphics[width=2.8 in]{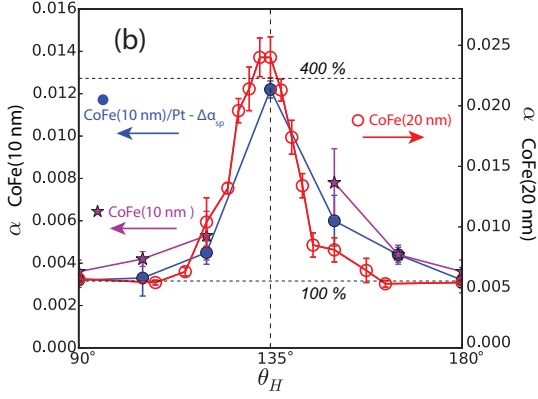}
 \caption{Renormalized damping and its anisotropy for CoFe(10 nm) and CoFe(20 nm), measured from spin-torque FMR and VNA FMR, respectively. For CoFe(20 nm)/Pt samples, $\Delta \alpha_{sp}$ has been subtracted from the measured damping.}
 \label{fig4}
\end{figure}
\indent In order to investigate the dominant mechanism for such a large Gilbert damping anisotropy, we perform temperature-dependent measurements of $\alpha$ and the resistivity $\rho$. Fig. \ref{fig5}(a) plots $\alpha$ as a function of $1/\rho$ for the CoFe(10 nm)/Pt and CoFe(20 nm) samples and for $H_B$ along the easy and hard axes. The dominant linear dependence reveals a major role of conductivitylike damping behavior. This is described by the breathing Fermi surface model for transition-metal ferromagnets, in which $\alpha$ can be expressed as \cite{kamberskyCJP1970,KamberskyPRB2007,GilmorePRL2007,FahnleJPD2008,GilmorePRB2010}:
\begin{equation}
\alpha \sim N(E_F)|\Gamma^-|^2 \tau
\label{eq1}
\end{equation}
where $N(E_F)$ is the density of state at the Fermi level, $\tau$ is the electron relaxation time and $\Gamma^-=\langle[\sigma^-, H_{so}]\rangle_{E=E_F}$ is the matrix for spin-flip scatterings induced by the SOC Hamiltonian $H_{so}$ near the Fermi surface \cite{KamberskyPRB2007,GilmorePRL2007}. Here $\tau$ is proportional to the conductivity ($1/\rho$) from the Drude model, with which Eq. (\ref{eq1}) gives rise to the behaviors shown in Fig. \ref{fig5}(a). \\
\indent For the origin of damping anisotropy, we first check the role of $N(E_F)$ by \textit{ab-initio} calculations for different ordered cubic supercells, which is shown in the Supplemental Materials \cite{supplement}. However, a negligible anisotropy in $N(E_F)$ is found for different magnetization orientations. This is consistent with the calculated anisotropy in Ref. \cite{ChenNPhys2018}, where less than 0.4\% change of $N(E_F)$ was obtained in ultrathin Fe films. The role of $\tau$ can also be excluded from the fact that the resistivity difference between the easy and hard axes is less than 2\% \cite{supplement}. Thus we deduce that the giant damping anisotropy of 400\% is due to the change of $|\Gamma^-|^2$, or the SOC, at different crystalline directions. In particular, unlike the single element Fe, disordered bcc Fe-Co alloy can possess atomic short-range order, which gives rise to local tetragonal crystal distortions due to the different lattice constants of Fe and Co \cite{razee1999,kota2012,TurekPRB2012}. Such local tetragonal distortions will preserve global cubic symmetry but can have large effects on the SOC. We emphasize that our CoFe samples, which did not experience annealing, preserve the random disorder. Our first principle calculations also confirm the role of local tetragonal distortions and its enhancement on SOC, see the Supplemental Materials for details \cite{supplement}.\\
\indent The anisotropy of the SOC in Co$_{50}$Fe$_{50}$ can be reflected by its AMR variation along different crystalline orientations. The AMR ratio can be defined as:
\begin{equation}
\text{AMR}(\theta_I) = {\rho_{\parallel}(\theta_I)\over\rho_\perp(\theta_I)}-1
\label{eq2}
\end{equation}
where $\rho_{\parallel}(\theta_I)$ and $\rho_{\perp}(\theta_I)$ are measured for the biasing field parallel and perpendicular to the current direction, respectively. The main contribution of AMR is the asymmetric $s$-$d$ electron scatterings where the $s$-orbitals are mixed with magnetization-containing $d$-orbitals due to SOC \cite{McGuireIEEE1975,PotterPRB1974}. Since both the damping and AMR originate from SOC and, more precisely, are proportional to the second order of SOC, a large damping anisotropy is expected to be accompanied by a large AMR anisotropy and vice versa. Furthermore, due to the fourfold symmetry, the AMR should be invariant when the current direction is rotated by 90 degrees, as the AMR is a function of $\theta_I$ as defined in Eq. (\ref{eq1}). Thus the damping and AMR should exhibit similar angular dependence on $\theta_H$ and $\theta_I$, respectively.\\
\begin{figure}[htb]
 \centering
 \includegraphics[width=2.8 in]{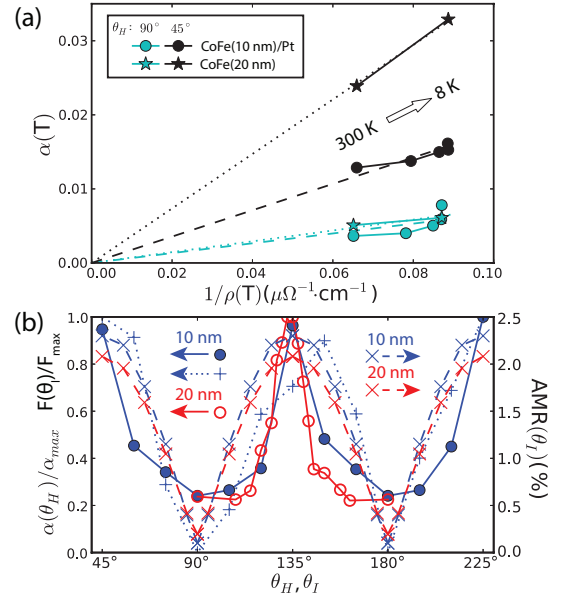}
 \caption{(a) $\alpha(T)$ as a function of $1/\rho(T)$. $T=8$ K, 30 K, 70 K, 150 K and 300 K for CoFe(10 nm)/Pt and $T=8$ K and 300 K for CoFe(20 nm). Dashed and dotted lines are guides to eyes. (b) Renormalized $\alpha(\theta_H)$ and $\text{AMR}(\theta_I)$ and $F(\theta_I)$ for CoFe(10 nm)/Pt and CoFe(20 nm). Circles, crosses and $+$ denote $\alpha$, AMR and $F$, respectively.}
 \label{fig5}
\end{figure}
\indent In Fig. \ref{fig5}(b) we compare renormalized $\alpha(\theta_H)$ with $\text{AMR}(\theta_I)$ for 10-nm and 20-nm CoFe samples, where the AMR values are measured from Hall bars with different $\theta_I$. The AMR ratio is maximized along $\langle 100 \rangle$ axes and minimized along $\langle 110 \rangle$ axes, with a large anisotropy by a factor of 10. This anisotropy is also shown by the integrated spin-torque FMR intensity for CoFe(10 nm)/Pt, defined as $F(\theta_I)=\Delta H_{1/2} V_{dc}^{max}$ \cite{LiuPRL2011,WeiZhangPRB2015} and plotted in Fig. \ref{fig5}(b). The large AMR anisotropy and its symmetry clearly coincide with the damping anisotropy measured in the same samples, which confirms our hypothesis of strong SOC anisotropy in CoFe. Thus we conclude that the damping anisotropy is dominated by the variation of SOC term in Eq. (\ref{eq1}). In parallel, we also compare $\alpha(\theta_H)$ and $\text{AMR}(\theta_I)$ for epitaxial Fe(10 nm) samples grown on GaAs substrates \cite{supplement}. Experimentally we find the anisotropy less is than 30\% for both damping and AMR, which helps to explain the presence of weak damping anisotropy in epitaxial Fe \cite{ChenNPhys2018}.\\
\indent We compare our results with prior theoretical works on damping anisotropy \cite{FahnleJPD2008,GilmorePRB2010}. First, despite their proportional relationship in Fig. \ref{fig5}(a), the giant anisotropy in $\alpha$ is not reflected in $1/\rho$. This is because the $s$-$d$ scattering, which dominates in the anisotropic AMR, only contributes a small portion to the total resistivity. Second, neither the anisotropy of damping nor AMR are sensitive to temperature. This is likely because the thermal excitations at room temperature ($\sim 0.025$ eV) are much smaller than the spin-orbit coupling ($\sim 0.1$ eV \cite{kamberskyCJP1970}). Third, the damping tensor has been expressed as a function of $\mathbf{M}$ and $d\mathbf{M}/dt$ \cite{GilmorePRB2010}. However in a fourfold-symmetry lattice and considering the large precession ellipticity, these two vectors are mostly perpendicular to each other, point towards equivalent crystalline directions, and contribute equivalently to the symmetry of damping anisotropy.\\
\indent In summary, we have experimentally demonstrated very large Gilbert damping anisotropy up to 400\% in epitaxial Co$_{50}$Fe$_{50}$ thin films which is due to their bulk, cubic crystalline anisotropy. We show that the damping anisotropy can be explained by the change of spin-orbit coupling within the breathing Fermi surface model, which can be probed by the corresponding AMR change. Our results provide new insights to the damping mechanism in metallic ferromagnets, which are important for optimizing dynamic properties of future magnetic devices.\\
\indent We are grateful for fruitful discussions with Bret Heinrich. W.Z. acknowledges supports from the U.S. National Science Foundation under Grants DMR-1808892, Michigan Space Grant Consortium and DOE Visiting Faculty Program. Work at Argonne, including transport measurements and theoretical modeling, was supported by the U.S. Department of Energy, Office of Science, Materials Science and Engineering Division. Work at Fudan, including thin film growth and fabrication, was supported by the Nat'l Key Basic Research Program (2015CB921401), Nat'l Key Research and Development Program (2016YFA0300703), NSFC (11734006,11474066,11434003), and the Program of Shanghai Academic Research Leader (17XD1400400) of China. O.O. and V.K. acknowledge supports from Bogazici University Research Fund (17B03D3), TUBITAK 2214/A and U.S. Department of State Fulbright Visiting Scholar Program.

\newpage

\onecolumngrid

\newpage

\Large{\textbf{Supplemental Materials:Giant anisotropy of Gilbert damping in epitaxial CoFe films}}\\
\normalsize
\textit{by} Yi Li, Fanlong Zeng, Steven S.-L. Zhang, Hyeondeok Shin, Hilal Saglam, Vedat Karakas, Ozhan Ozatay, John E. Pearson, Olle G. Heinonen, Yizheng Wu, Axel Hoffmann and Wei Zhang
\newline

\renewcommand{\theequation}{S-\arabic{equation}}
\setcounter{equation}{0}  
\renewcommand{\thefigure}{S-\arabic{figure}}
\setcounter{figure}{0}  

\subsection{Crystallographic quality of Co$_{50}$Fe$_{50}$ films}

\begin{figure}[htb]
 \centering
 \includegraphics[width=6.0 in]{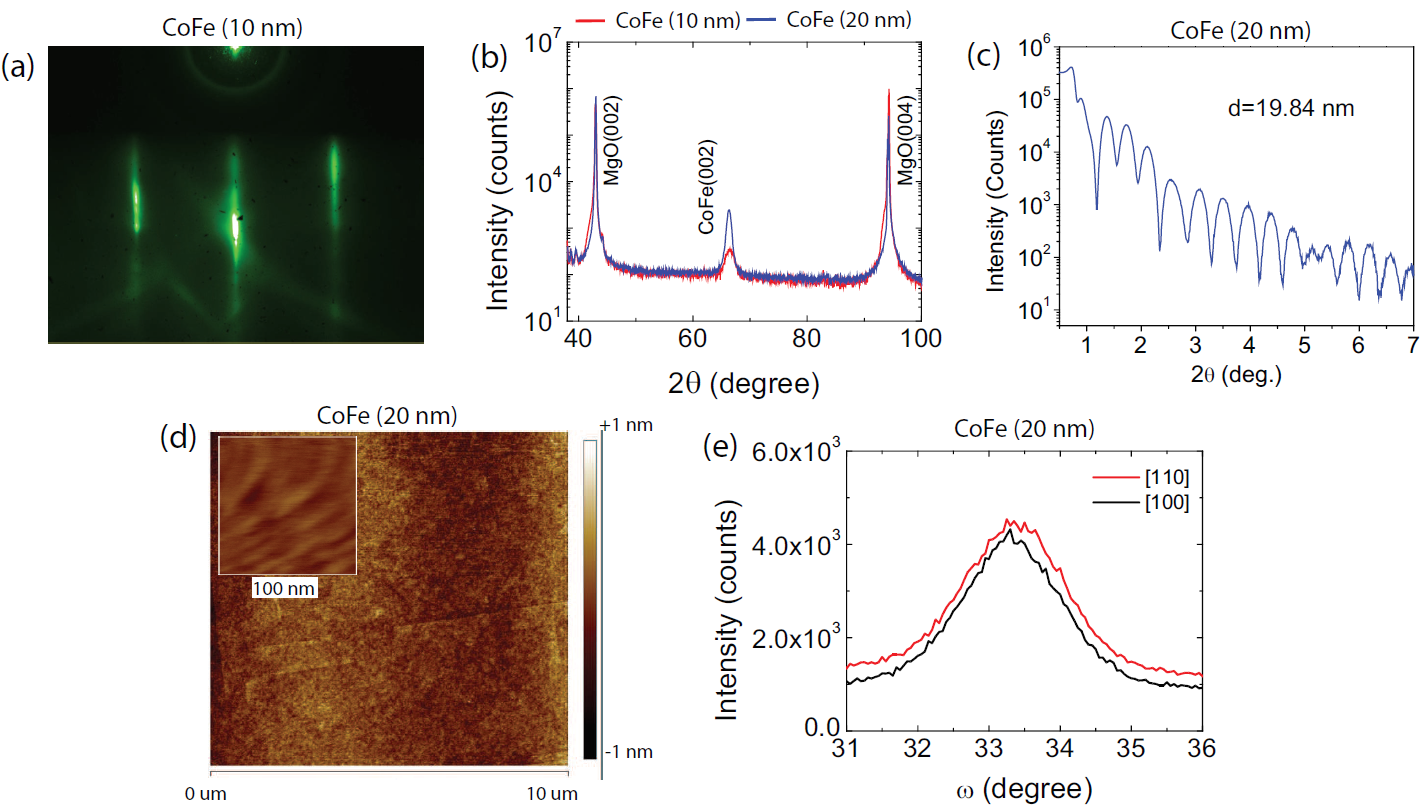}
 \caption{Crystallographic characterization results of CoFe films. (a) RHEED pattern of the CoFe(10 nm) film. (b) XRD of the CoFe(10 nm) and (20 nm) films. (c) X-ray reflectometry measured for the CoFe(20 nm) film. (d) AFM scans of the CoFe(20 nm) film. (e) Rocking curves of the CoFe(20 nm) film for [100] and [110] rotating axes.}
 \label{sm0}
\end{figure}

Fig. \ref{sm0} shows the crystallographic characterization for the epitaxial CoFe samples. Reflection high-energy electron diffraction (RHEED) shows very clear and sharp patterns which shows high quality of the epitaxal films. X-ray diffraction (XRD) yields clear CoFe(002) peaks at $2\theta=66.5^\circ$. X-ray reflectometry scan of the CoFe (20 nm) film shows a good periodic pattern and the fit gives a total thickness of 19.84 nm. Atomic-force microscopy (AFM) scans for 10 $\mu$m $\times$ 10 $\mu$m and 100 nm $\times$ 100 nm scales show smooth surface with a roughness of 0.1 nm. Lastly XRD rocking curves for [100] and [110] rotating axes show a consistent linewidth of 1.45$^\circ$, which indicates isotropic mosaicity of the CoFe films.

As a result of the crystallographic characterizations, we believe our MBE-grown CoFe samples are epitaxial, have smooth surfaces and exhibit excellent crystalline quality. Moreover, we can exclude the source of inhomogeneity from misorientation of crystallities (mosaicity) due to isotropic rocking curves. This means the inhomogeneous FMR linewidth broadening is isotropic, as is consistent with the experiments.\\

\subsection{Device geometries for Spin-torque FMR and VNA FMR measurements.}

\begin{figure}[htb]
 \centering
 \includegraphics[width=6.0 in]{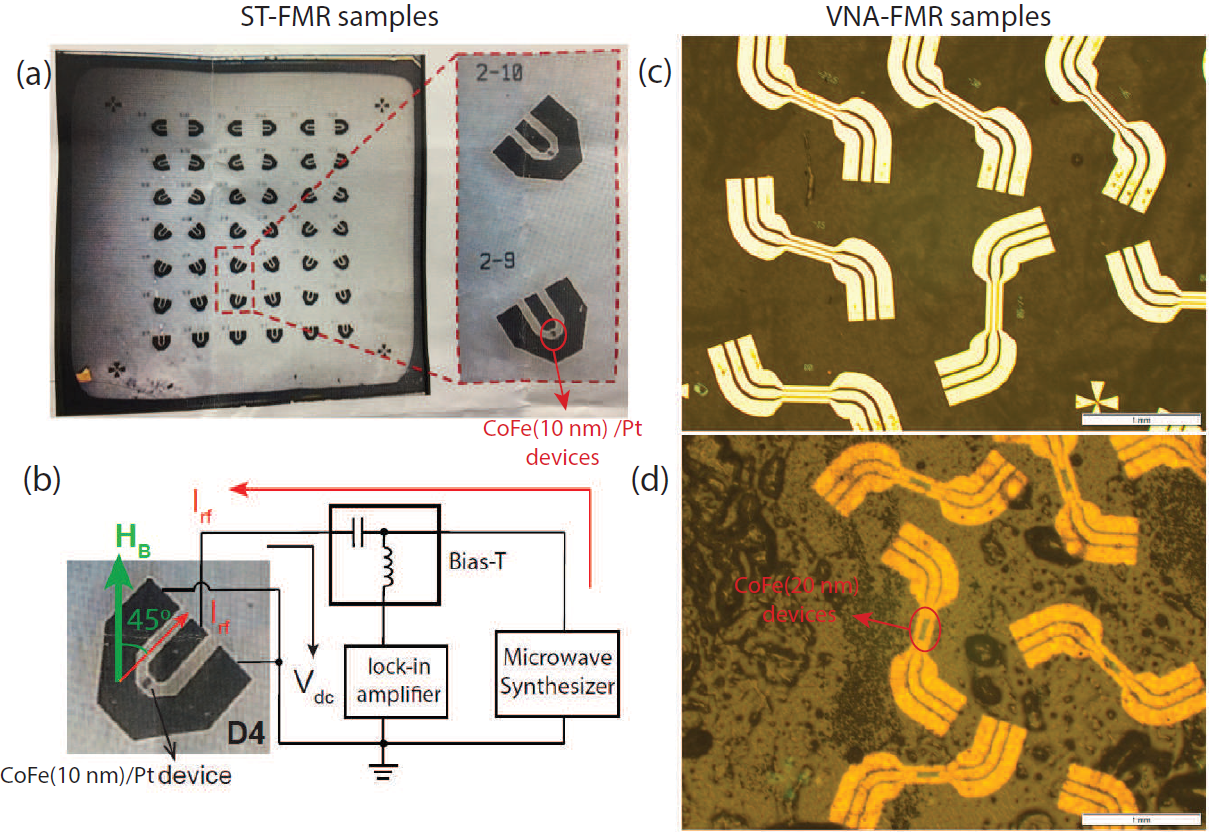}
 \caption{(a) Spin-torque FMR devices of CoFe(10 nm)/Pt samples. (b) Illustration of the Spin-torque FMR circuit. (c) Front and (d) back view of the VNA FMR devices for CoFe(20 nm) samples.}
 \label{sm0_5}
\end{figure}

Fig. \ref{sm0_5} shows the device geometry for Spin-torque FMR and VNA FMR measurements. For spin-torque FMR, we have prepared CoFe(10 nm)/Pt, CoFe(10 nm) and Fe(10 nm) devices. A second CoFe(10 nm)/Pt sample is also prepared for rotating-field measurements. For VNA FMR, we have prepared CoFe(20 nm) samples. All the CoFe films are grown on MgO(100) substrates; the Fe film is grown on a GaAs(100) substrate. Au (100 nm) coplanar waveguides are subsequently fabricated on top of all devices. For VNA FMR samples, an additional SiO$_2$(100 nm) is deposited between CoFe and Au for electric isolation. The CoFe(20 nm) bars is only visible from the back view in Fig. \ref{sm0_5}(d).

\subsection{Spin-torque FMR lineshapes}

\begin{figure}[htb]
 \centering
 \includegraphics[width=6.0 in]{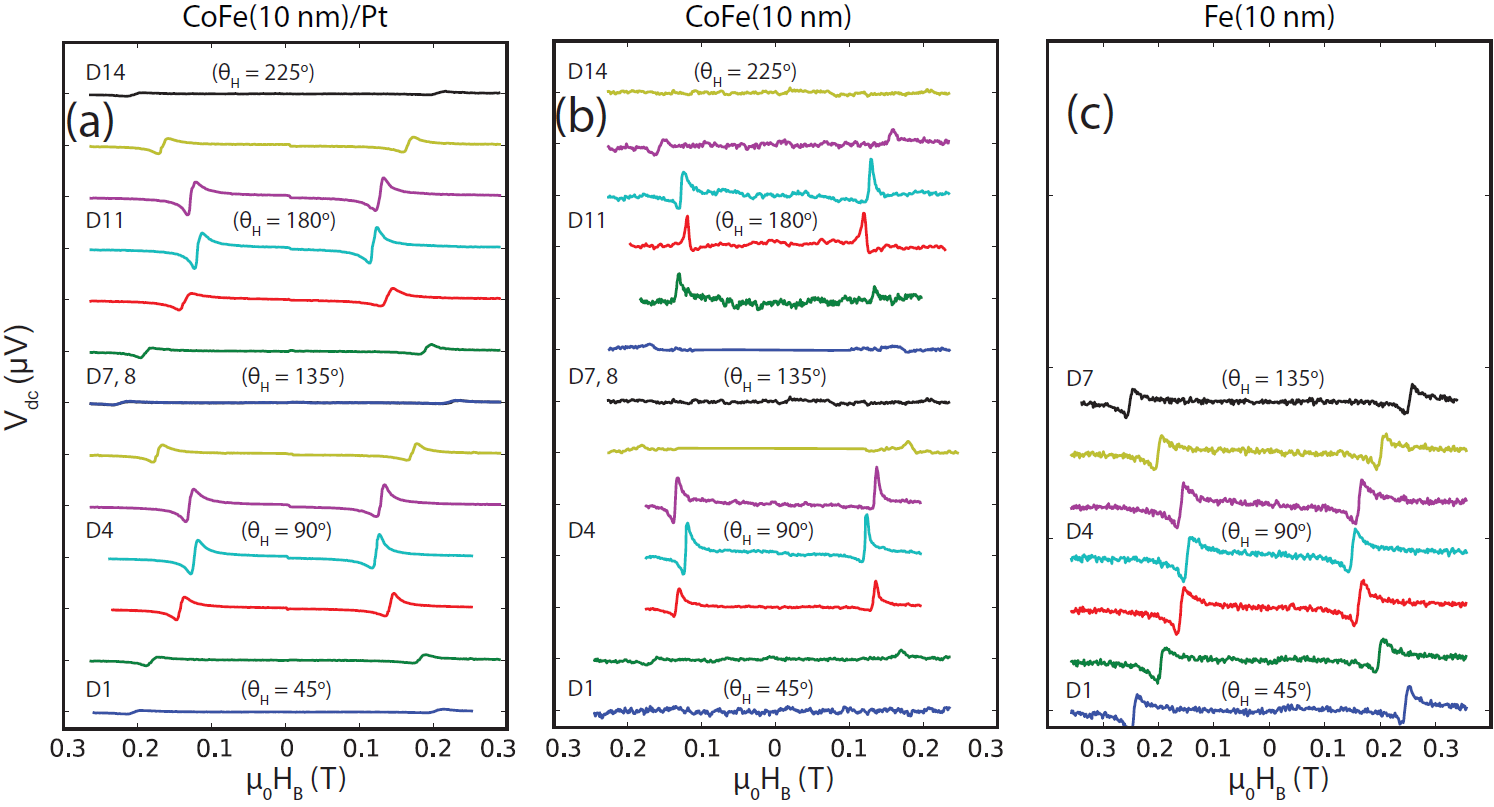}
 \caption{Spin-torque FMR lineshapes of (a) CoFe(10 nm)/Pt, (b) CoFe(10 nm) and (c) Fe(10 nm) devices measured at $\omega/2\pi=20$ GHz. $\theta_H-\theta_I$ is fixed to 45$^\circ$.}
 \label{sm1}
\end{figure}

Figure \ref{sm1} shows the full lineshapes of (a) CoFe(10 nm)/Pt(6 nm), (b) CoFe(10 nm) and (c) Fe(10 nm) devices measured at $\omega/2\pi=20$ GHz. The Fe films were deposited on GaAs substrates by MBE growth. (a) and (b) are used to extract the resonance fields and linewidths in Figs. 2(a) and 3(a) of the main text. (c) is used to examine the correlation between damping anisotropy and AMR anisotropy.

\subsection{Spin-torque FMR linewidths as a function of frequency for CoFe(10 nm) devices.}

\begin{figure}[htb]
 \centering
 \includegraphics[width=4.5 in]{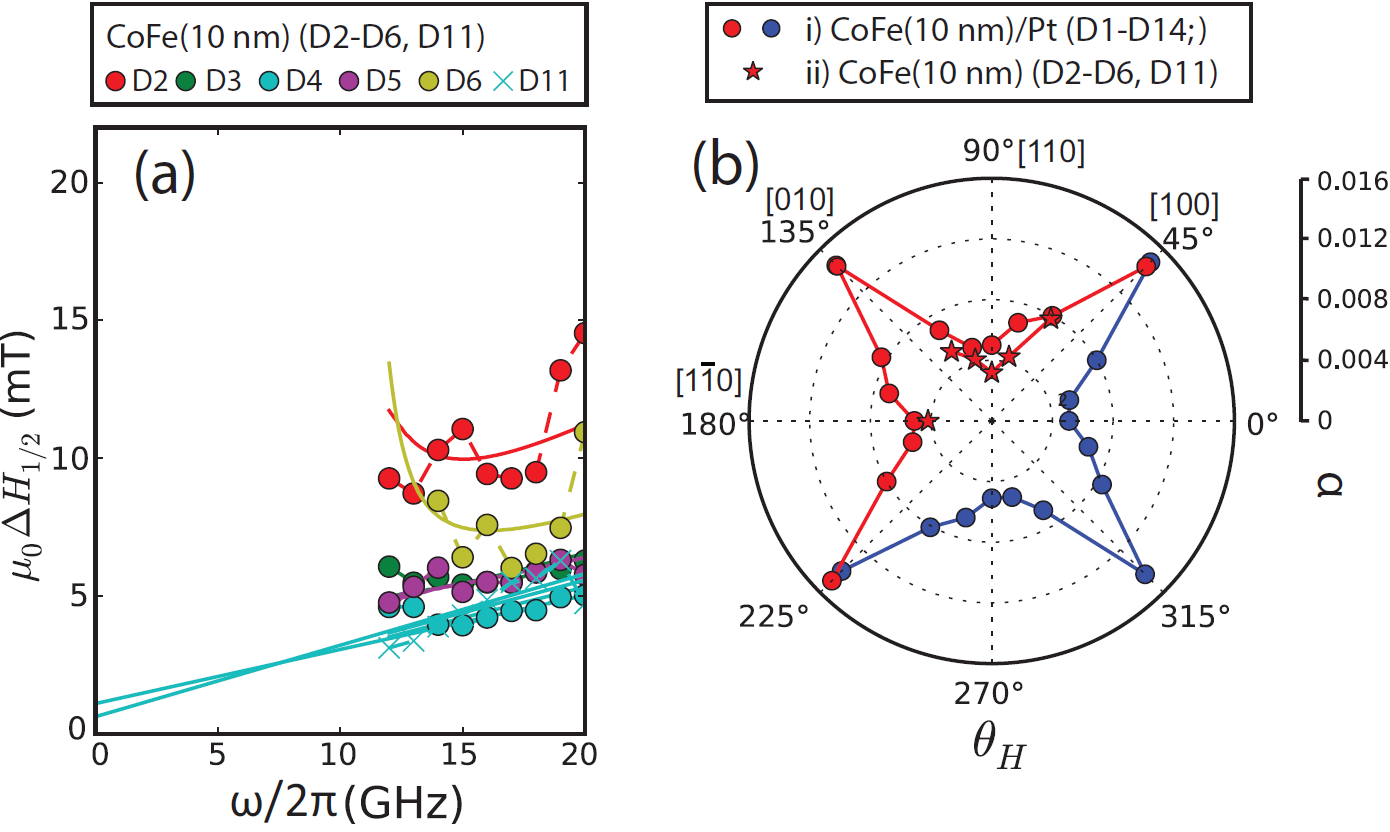}
 \caption{(a)$\mu_0\Delta H_{1/2}$ as a function of frequency for CoFe(10 nm) devices. Solid lines and curves are the fits to the experiments. $\theta_H-\theta_I$ is fixed to 45$^\circ$. (b) $\alpha$ as a function of $\theta_H$ for CoFe(10 nm)/Pt and CoFe(10 nm) devices.}
 \label{sm2}
\end{figure}

Figure \ref{sm2}(a) shows the spin-torque FMR linewidths for CoFe(10 nm) devices. Because there is no spin torque injection from Pt layer, the FMR signals are much weaker than CoFe(10 nm)/Pt and the extracted linewidths are more noisy. The excitation of the dynamics is due to the magnon charge pumping effect \cite{ciccarelliNNano2015} or inhomogeneities of the Oersted fields. No signal is measured for the rf current flowing along the easy axis (magnetic field along the hard axis, see Fig. \ref{sm1}b), because of the negligible AMR ratio.

Figure \ref{sm2}(b) shows the angular dependence of the extracted Gilbert damping for CoFe(10 nm)/Pt and CoFe(10 nm). The former is extracted from Fig. 3(b) of the main text. The latter is extracted from Fig. \ref{sm2}(a). The blue data points for CoFe(10 nm)/Pt are obtained from the resonances at negative biasing fields. Those data are used in Fig. 4 of the main text.\\
\

\subsection{Spin-torque FMR for CoFe(10 nm)/Pt up to 39 GHz.}

\begin{figure}[htb]
 \centering
 \includegraphics[width=5.5 in]{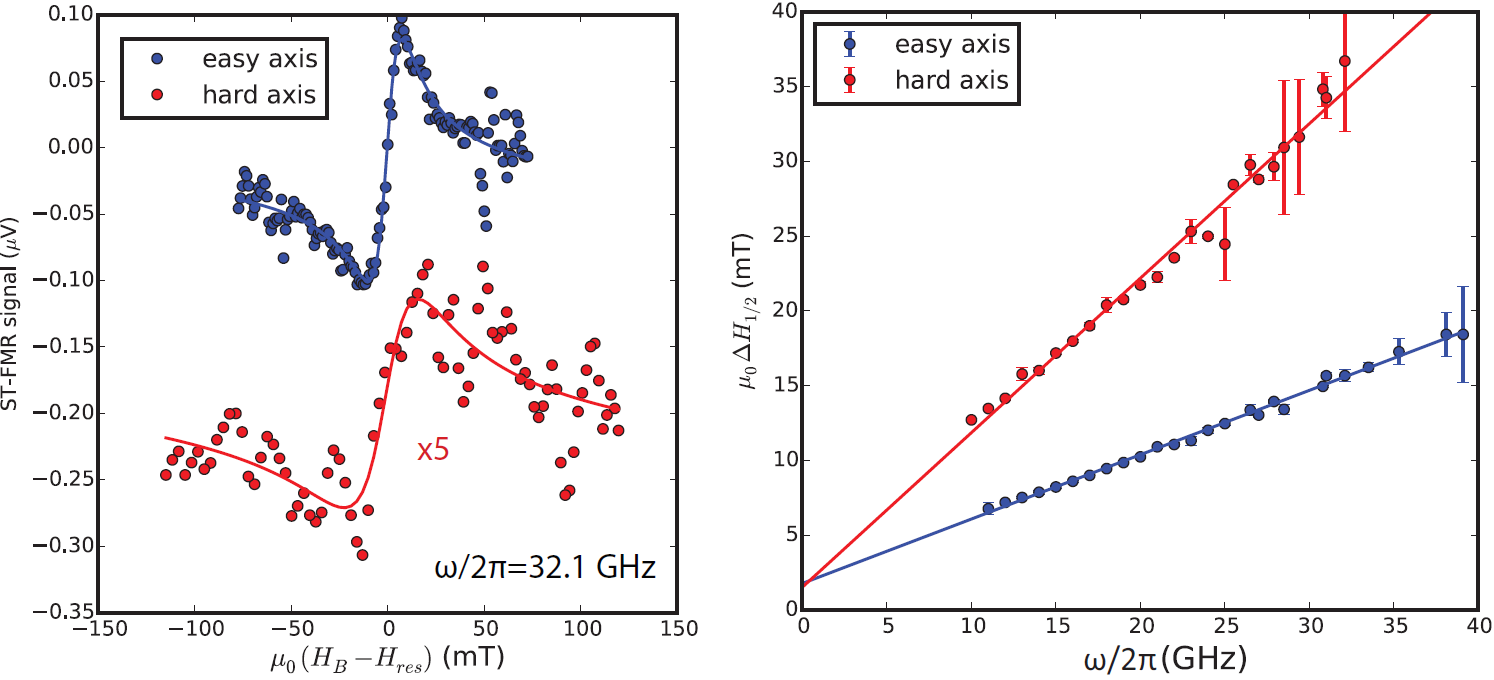}
 \caption{High-frequency ST-FMR measurement of i) CoFe(10 nm)/Pt for the biasing field along the easy axis ($\theta_H=90^\circ$) and hard axis ($\theta_H=45^\circ$). Left: lineshapes of ST-FMR at $\omega/2\pi=32.1$ GHz. Right: linewidth as a function of frequency. Lines are linear fits to the data by setting both $\alpha$ and $\Delta H_0$ as free parameters.}
 \label{hf_STFMR}
\end{figure}

Fig. \ref{hf_STFMR} shows the spin-torque FMR lineshapes and linewidths up to 39 GHz for CoFe(10 nm)/Pt devices along the easy and hard axes ($\theta_H = 90^\circ and 45^\circ$). At $\omega/2\pi=32.1$ GHz (Fig. \ref{hf_STFMR}a), the spin-torque FMR amplitude is 0.1 $\mu$V for the easy axis and 0.02 $\mu$V for the hard axis. 10 seconds of time constant is used to obtained the signals. Throughout the frequency range, linewidths demonstrate good linear dependence on frequency as shown Fig. \ref{hf_STFMR}(b). For the hard axis the signal has reached the noise bottom limit at 32.1 GHz. For the easy axis the noise bottom limit is reached at 39 GHz. The two linear fits yield $\alpha=0.0063$ and $\mu_0\Delta H_0=1.8$ mT for the easy axis and $\alpha=0.00153$ and $\mu_0\Delta H_0=1.5$ mT for the hard axis. The two damping parameters are close to the values obtained below 20 GHz in the main text. Also the inhomogeneous linewidth $\mu_0\Delta H_0$ nicely match between easy and hard axes.

\subsection{Low-temperature FMR linewidths and dampings for CoFe(10 nm)/Pt and CoFe(20 nm).}

\begin{figure}[htb]
 \centering
 \includegraphics[width=6.0 in]{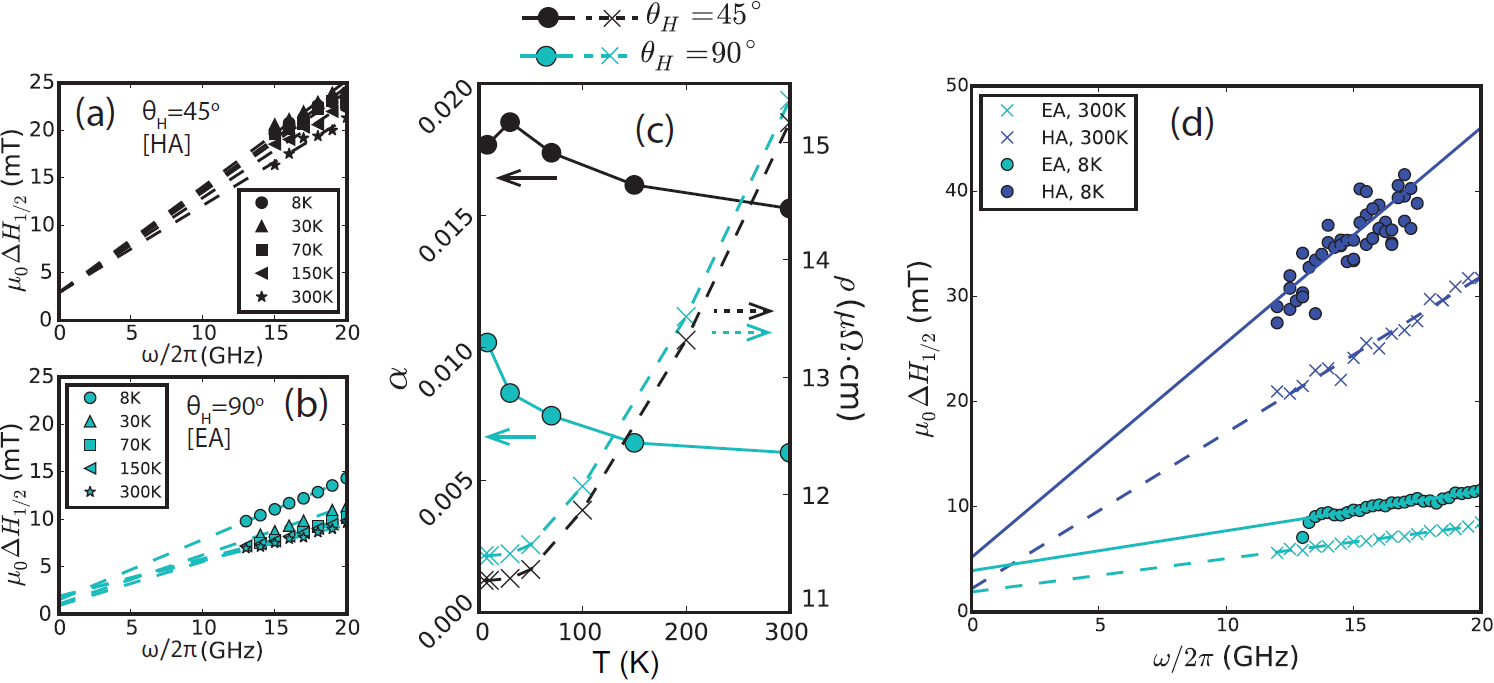}
 \caption{(a-b)$\mu_0\Delta H_{1/2}$ as a function of frequency for CoFe(10 nm)/Pt devices at different temperatures. (c) Extracted damping at different temperatures, same as in Fig. 4 of the main text.
 }
 \label{sm3}
\end{figure}

Figure \ref{sm3} shows the frequency dependence of linewidths for extracting temperature-dependent Gilbert damping in Fig. 5(a) of the main text.

For CoFe(10 nm)/Pt samples, we plot both $\alpha$ and resistivity $\rho$ measured at different temperatures in Fig. \ref{sm3}(c). The measurements of $\rho$ were conducted with a biasing magnetic field of 1 Tesla parallel to the current direction, so that the AMR influence is excluded. Also the resistivity variation between the easy and hard axes is very small, about 1\%, which is much smaller than the damping anisotropy.

We have also conducted the low-temperature VNA FMR of the new CoFe(20 nm) samples at 8 K, in addition to the room-temperature measurements. The linewidths data are shown in Fig. \ref{sm3}(d) for both easy and hard axes. The extracted damping are: $\alpha=0.0054$ (EA, 300 K), 0.0061 (EA, 8 K), 0.0240 (HA, 300 K) and 0.0329 (HA, 8 K). Those values are used in Fig. 4(b) and Fig. 5(a) of the main text.

For CoFe(10 nm) the damping anisotropy decreases from 380 \% at 300 K to 273 \% at 30 K by taking out the spin pumping damping enhancement (an unexpected reduction of $alpha$ happens at 8 K for the hard axis). For CoFe(20 nm) the damping anisotropy increases from 440 \% at 300 K to 540 \% at 8 K. Thus a clear variation trend of damping anisotropy in CoFe films remains to be explored.

\subsection{First-principle calculation of $N(E_F)$ anisotropy for Co$_{50}$Fe$_{50}$}

\begin{figure}[htb]
 \centering
 \includegraphics[width=4.0 in]{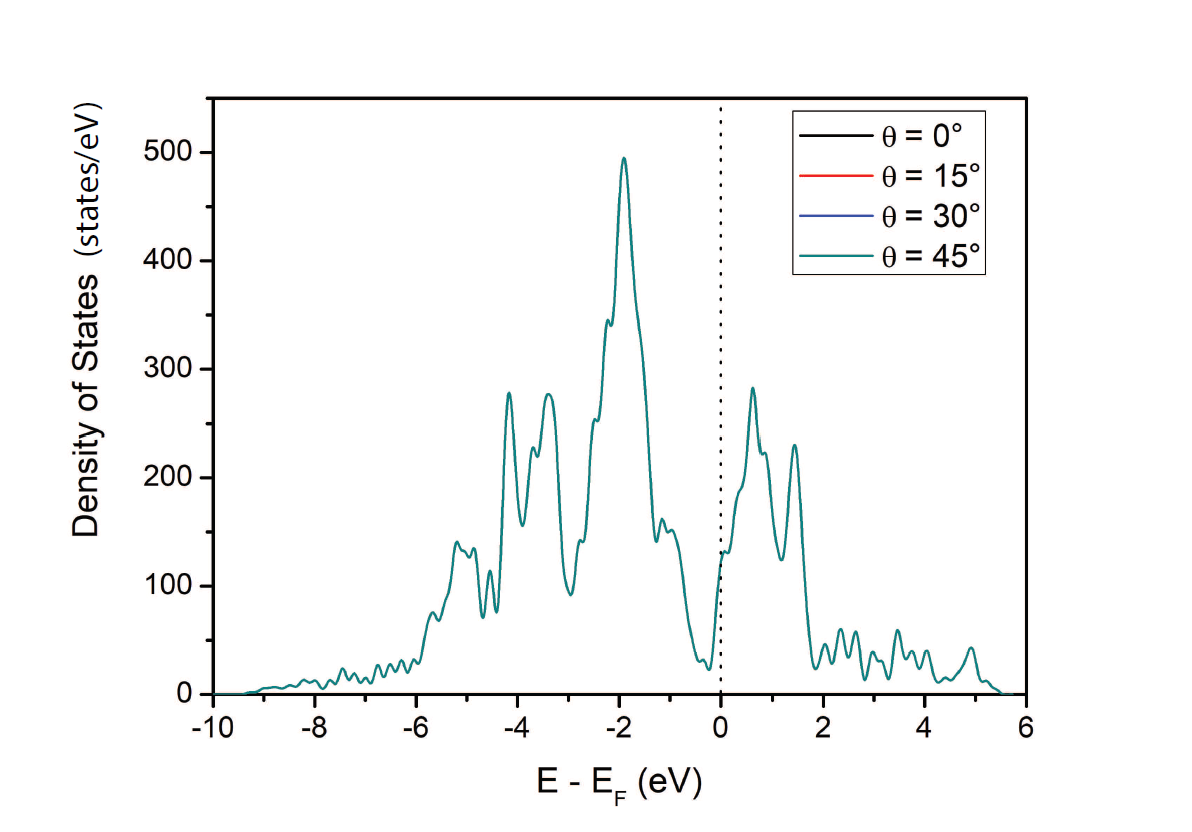}
 \caption{Density of states as a function of energy. $E_F$ is the Fermi level.}
 \label{sm4}
\end{figure}

First-principle calculations were done using QUANTUM ESPRESSO for a cubic lattice of Co$_{50}$Fe$_{50}$ of CsCl, Zintl and random alloy structures. Supercells consisting of $4\times4\times4$ unit cells were considered with a total of 128 atoms (64 cobalt and 64 iron atoms). The calculations were done using plane-wave basis set with a 180 Ry kinetic energy cut-off and 1440 Ry density cut-off. For both Co and Fe atoms, fully  relativistic PAW pseudopotentials were used. Figure \ref{sm4} shows the density of states (DOS) of the CsCl form for different magnetization orientations $\theta$ in the $xy$-plane. Clearly, DOS exhibits no anisotropy ($<0.1$\% variation at $E=E_F$). No anisotropy was found in the Zintl form, either. Thus, we conclude that the Gilbert damping anisotropy in Co$_{50}$Fe$_{50}$ cannot be caused by a variation of $N(E_F)$ with respect to magnetization direction in ideal ordered structures.

\subsection{SOC induced by atomic short-range order (ASRO)}

In our experiment, because the Co$_{50}$Fe$_{50}$ films were grown by MBE at low temperatures, they do not form the ordered bcc B2 structure but instead exhibit compositional disorder. Transition metal alloys such as CoPt, NiFe, and CoFe tend to exhibit ASRO \cite{razee1999,kota2012,TurekPRB2012}. The ASRO in CoFe is likely to give rise to local tetragonal distortions because of the different lattice constants of bcc Fe and (metastable) bcc Co at 2.856~{\AA} and 2.82~{\AA}, respectively. Such local tetragonal distortions will preserve global cubic (or four-fold in-plane) symmetry, but can have large effects on the SOC, with concomitant effect on spin-orbit induced magnetization damping. For example, first-principle calculations using the coherent-potential approximation for the  substitutionally disordered system shows that a tetragonal distortion of 10\% in the ratio of the tetragonal axes $a$ and $c$ gives rise to an magnetocrystalline anisotropy energy (MAE) density \cite{razee1999,kota2012} of about 1 MJ/m$^3$. These results are consistent with our observed MAE in Co$_{50}$Fe$_{50}$.

\begin{figure}[htb]
 \centering
 \includegraphics[width=6.0 in]{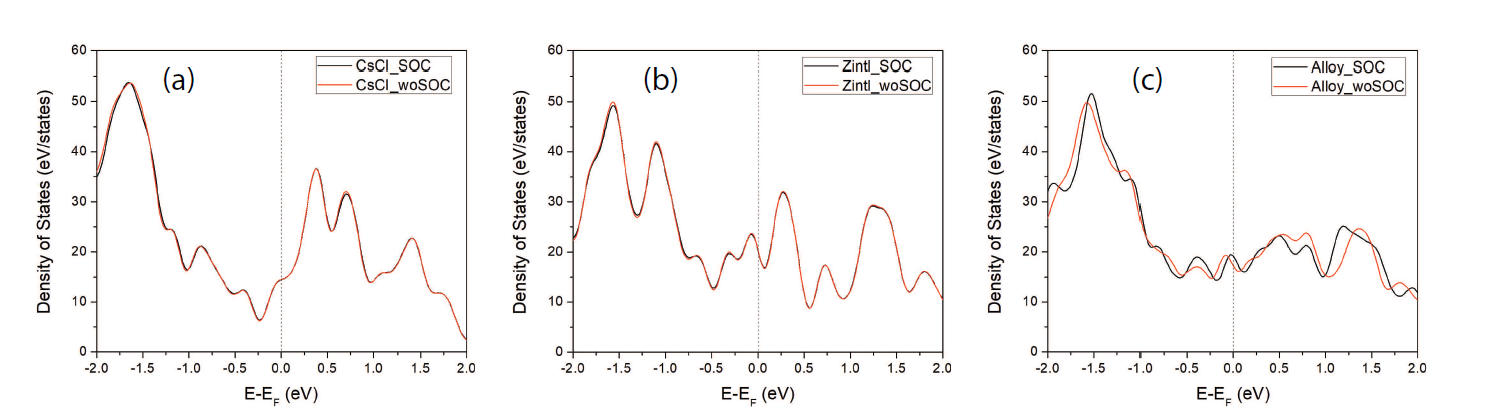}
 \caption{Density of states (DOS) for (a) CsCl, (b) Zintl, and (c) alloy form of CoFe with SOC (black solid) and without SOC (red solid).}
 \label{sm5}
\end{figure}

To confirm this mechanism, we performed DFT-LDA calculations on 50:50 CoFe supercells consisting of a total of 16 atoms for CsCl, zintl, and random alloy structures; in the random alloy supercell, Co or Fe atoms randomly occupied the atomic positions in the supercell. Note that all CoFe geometries are fully relaxed, including supercell lattice vectors.

\begin{figure}[htb]
 \centering
 \includegraphics[width=6.0 in]{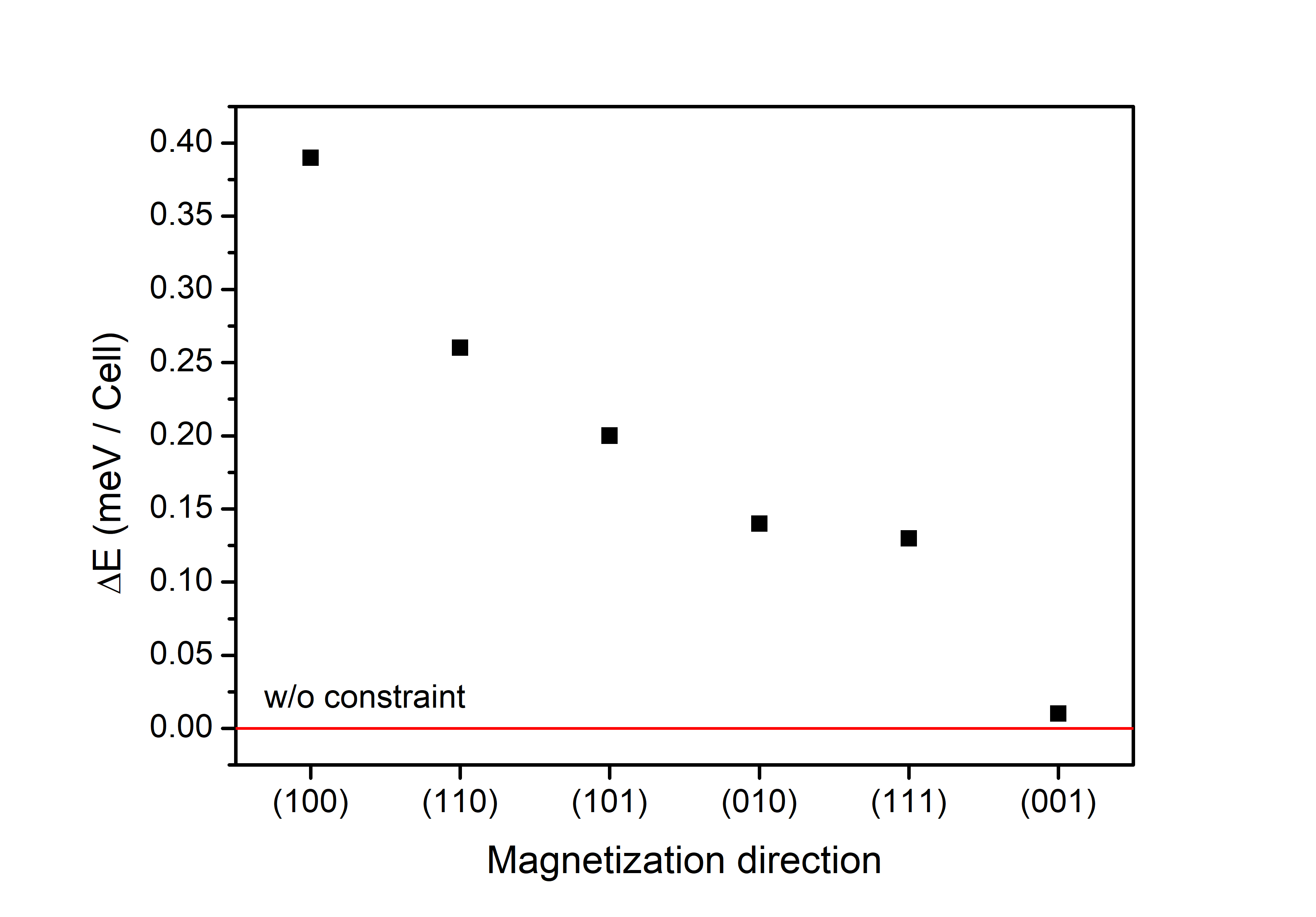}
 \caption{Change in total energy (per supercell) of the alloy structure as function of the magnetization direction.}
 \label{fig:SI_energy}
\end{figure}

\begin{enumerate}
	\item Structural relaxation including spin-orbit coupling (SOC) shows local tetragonal distortions for random alloy supercell. Among the three different CoFe phases, tetragonal c/a ratio for the supercell in optimized geometry is largest (1.003) in the random alloy supercell with SOC, which means local tetragonal distortions are more dominant in random alloy compared to CsCl and Zintl structures. [c/a values : CsCl (0.999), Zintl (0.999), Alloy (1.003)]. In addition, the alloy system exhibited local distortions of Co and Fe position relative to their ideal positions. In contrast, in CsCl and Zintl structures the Co and Fe atoms exhibited almost imperceptible distortions. Table~\ref{table:SI1} shows the relaxed atomic positions in the alloys structure in units of the lattice vectors. In the ideal (unrelaxed) system, the positions are all at multiples of 0.25; the relaxed CsCl and Zintl structures no deviations from these positions larger than 1 part in $10^6$
    \item SOC changes the density of states (DOS) at the Fermi energy, notably for the random alloy but {\em not} for the CsCl and Zintl structures. Figure~\ref{sm5} shows DOS for (a) CsCl, (b) Zintl, and (c) random alloy structure with SOC (black lines) and without it red lines). We can see significant DOS difference for the random alloy supercell with SOC where tetragonal distortions occurred, while almost no changes are observed in the CsCl and Zintl structures.
    \item The local distortions in the alloy structure furthermore gave rise to an energy anisotropy with respect to the magnetization direction. The energy (including SOC) of the relaxed alloy structure for different directions of the magnetization is shown in Fig.~\ref{fig:SI_energy}. While the supercell was rather small, because of the computational expense in relaxing the structure with SOC, so that no self-averaging can be inferred, the figure demonstrates an induced magnetic anisotropy that arises from the SOC and local distortions. No magnetic anisotropy was discernible in the CsCl and Zintl structures.
\end{enumerate}
\begin{table}
\label{table:SI1}
\caption{Relaxed atomic positions (including SOC) of the alloy structure. In the ideal CsCl or Zintl structures, the atomic positions are all multiples of 0.25 in units of the lattice vector components.}
\begin{ruledtabular}
\begin{tabular}{cccc}
Atom & x-position & y-position & z-position \\ \hline
Co & 0.003783083 & 0.000000000 & 0.000000000\\
Fe & -0.001339230 & 0.000000000 &  0.500000000\\
Fe & -0.002327721 &  0.500000000 &  0.000000000\\
Fe & 0.002079922 &  0.500000000  & 0.500000000\\
Fe & 0.502327721 &  0.000000000 & 0.000000000\\
Fe & 0.497920078 &  0.000000000 &  0.500000000\\
Co & 0.496216917 &  0.500000000 & 0.000000000\\
Fe & 0.501339230 &  0.500000000 &  0.500000000\\
Co & 0.250000000 &  0.250000000 &  0.254117992\\
Fe & 0.250000000 &  0.250000000 &  0.752628048\\
Fe & 0.250000000 &  0.750000000 &  0.247371952\\
Co & 0.250000000 &  0.750000000 &  0.745882008\\
Co & 0.750000000 &  0.250000000 &  0.250415490\\
Co & 0.750000000 &  0.250000000 &  0.746688258\\
Co & 0.750000000 &  0.750000000 &  0.253311742\\
Co & 0.750000000 &  0.750000000 &  0.749584510\\
\end{tabular}
\end{ruledtabular}
\end{table}

As a result from the DFT calculation, we attribute the large SOC effect in damping anisotropy of Co$_{50}$Fe$_{50}$ to local tetragonal distortions in disordered Co and Fe alloys. These distortions give rise to SOC-induced changes of DOS at the Fermi level, as well as magnetic anisotropy energy with respect to the crystallographic axes.

\bibliographystyle{ieeetr}

\end{document}